\documentclass{PoS}
\usepackage{graphicx}

\title{Exploring the powering source of the TeV X-ray binary LS 5039}

\ShortTitle{Exploring the powering source of the TeV X-ray binary LS 5039}

\author{\speaker{Javier Mold\'on}, Marc Rib\'o, Josep M. Paredes\\
Departament d'Astronomia i Meteorologia and Institut de Ci\`encies del Cosmos (ICC), Universitat de Barcelona\\
Mart\'{\i} i Franqu\`es, 1\\
08028 Barcelona, Spain\\
E-mail: \email{jmoldon@am.ub.es}, \email{mribo@am.ub.es}, \email{jmparedes@ub.edu}}

\author{Josep Mart\'i\\
        Departamento de F\'isica (EPS), Universidad de Ja\'en\\
        E-mail: \email{jmarti@ujaen.es}}
        
\author{Maria Massi\\
        Max Planck Institut für Radioastronomie\\
        E-mail: \email{mmassi@mpifr-bonn.mpg.de}}
   
\abstract{LS 5039 is one of the four TeV emitting X-ray binaries detected up to
now. The powering source of its multi-wavelength emission can be accretion in a
microquasar scenario or wind interaction in a young non-accreting pulsar
scenario. These two scenarios predict different morphologic and peak position
changes along the orbital cycle of 3.9 days, which can be tested at
milliarcsecond scales using VLBI techniques. Here we present a campaign of
5~GHz VLBA observations conducted in June 2000 (2 runs five days apart). The
results show a core component with a constant flux density, and a fast change
in the morphology and the position angle of the elongated extended emission,
but maintaining a stable flux density. These results are difficult to fit
comfortably within a microquasar scenario, whereas they appear to be compatible
with the predicted behavior for a non-accreting pulsar.}

\FullConference{The 9th European VLBI Network Symposium on The role of VLBI in the Golden Age for Radio Astronomy and EVN Users Meeting\\
		 September 23-26, 2008\\
		 Bologna, Italy}

\begin{document}

\section{Introduction} \label{introduction}

The very high energy emission of gamma-ray binaries (see \cite{aharonian05a,
aharonian05b, albert06, albert07}) is basically interpreted as the result of
inverse Compton upscattering of stellar UV photons by relativistic electrons.
Two excluding scenarios have been proposed to explain the acceleration
mechanism that powers the electrons. In the first one electrons are accelerated
in the jets of a microquasar powered by accretion (see \cite{bosch06},
\cite{paredes06}, and \cite{romero03}). In the second one they are accelerated
in the shock between the relativistic wind of a non-accreting pulsar and the
wind of the stellar companion (see \cite{maraschi81}, \cite{dubus06}, and
\cite{sierpowska07}).

LS~5039 is a gamma-ray binary situated at $2.5\pm0.5$~kpc, composed of a bright
young star and a compact object. The mass of the compact object depends on the
inclination of the orbit, which can be between 11 and 75$^{\circ}$, and is in
the range 1.5--8~${M}_{\odot}$. The compact object can thus be a black hole or
a neutron star \cite{casares05}.

\section{Testing the possible scenarios at mas scales and previous radio observations} \label{scenarios}

The expected behavior of the radio emission at mas scales is different in each
scenario. In the microquasar scenario we have a central core with extended
jet-like radio emission that displays a flux and distance asymmetry produced by
projection effects and the Doppler boosting \cite{mirabel99}, \cite{fender06}.
The direction of the jets can display long-term precession. The jets can be
bent or disrupted due to the interaction with the dense stellar wind of the
bright companion \cite{perucho08}. In the non-accreting pulsar scenario the
shocked material is contained in a 'bow' shaped nebula extending away from the
stellar companion that follows an elliptical path during the orbital cycle. At
mas scales, it is expected that the direction of the extended emission changes
with the pulsar's orbital motion. The peak of the radio emission, located at a
few AU behind the pulsar, should follow an elliptic orbit $\sim$10 times bigger
than the size of the orbit of the system \cite{dubus06}.

The unresolved radio emission of LS~5039 is persistent, non-thermal, and
variable, although no strong radio outburst or periodic variability have been
detected so far \cite{ribo99,clark01}. The source spectral index is
$\alpha=-0.46\pm0.01$ \cite{marti98}. No radio pulses have been detected at
1.4~GHz \cite{morris02}. However, free-free absorption with the stellar wind
may prevent the detection of pulsations, and observations at higher frequencies
are required. Observations at 5~GHz were performed using the VLBA and the VLA in its phased
array mode in 1999 \cite{paredes00}. The orbital phase of the system was  in
the range 0.12--0.15. The final synthesis map shows two-sided extended emission
emerging from a central core with a flux density of 12.8~mJy, and aligned
towards a direction with Position Angle (PA) $=125^{\circ}$, measured from
north to east. The source extends over 6~mas on the plane of the sky. The
morphology of the source was studied at larger scales with the EVN and MERLIN,
also showing bipolar extended emission emanating from a central core (see
Paredes et al.\ \cite{paredes02}).

\section{VLBA observations in June 2000} \label{observations}

We observed LS~5039 with the Very Long Baseline Array (VLBA) and the Very Large
Array (VLA), of the National Radio Astronomy Observatory (NRAO), at 5~GHz
frequency on 2000 June 3 and 8. The two 8-hour observing sessions, hereafter
run~A and run~B, took place at orbital phases in the range 0.71--0.79 and
0.43--0.51, respectively. The observations were performed switching between the
target source LS~5039 and the phase reference calibrator J1825$-$1718. The
correlation position of J1825$-$1718 was shifted by $\Delta\alpha=+57.0$~mas
and $\Delta\delta=+21.4$~mas from the accurate position provided by later
observations from the joint NASA/USNO/NRAO geodetic/astrometric program. In the
case of LS~5039, due to its proper motion, the source was found to be
$\Delta\alpha=+7.6$~mas and $\Delta\delta=-16.2$~mas away from the correlated
position. See details in \cite{ribo08}.

The post-correlation data reduction was performed using the Astronomical Image
Processing System ({\sc aips}). The VLA position had to be corrected by 0.43~m,
according to later geodetic measurements. The positions of J1825$-$1718 and
LS~5039 were corrected using the task {\sc clcor}. As recommended for
phase-referencing experiments, we applied ionospheric and Earth Orientation
Parameters corrections to the visibility data using {\sc clcor}. The fringe
fitting ({\sc fring}) of the residual delays and fringe rates was performed for
all the sources. Fringes for 15 and 25\% of the baselines were missing for the
target source LS~5039 and for the astrometric check source J1837$-$1532,
respectively. Typical data inspection and flagging were performed. An
independent reduction of the VLA data was performed using standard procedures
within {\sc aips}.

\section{Results} \label{results}

The VLA data correlated as an independent interferometer were compatible with a
point-like source for the obtained synthesized beam of
$5.5^{\prime\prime}\times3.7^{\prime\prime}$ in PA $\sim-4^\circ$. We measured
the flux density of the source every 30 minutes and obtained a mean of 29.4~mJy
with a standard deviation of $\sigma$=1.1~mJy during the 8 hours of run~A, and
28.4~mJy with $\sigma$=0.7~mJy for run~B.

\begin{figure}[t!]
\center
\includegraphics[angle=0,scale=0.64]{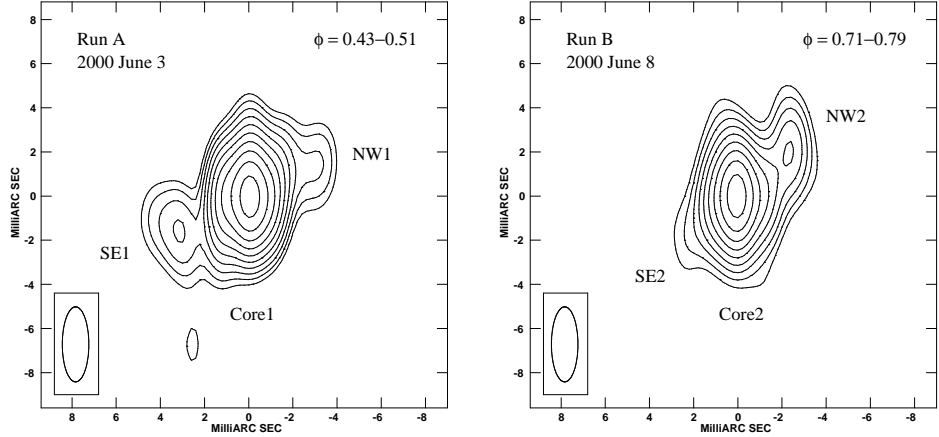}
\caption{VLBA+phased VLA self-calibrated images of LS~5039 at 5~GHz obtained on
2000 June 3 (left) and 8 (right). North is up and east is to the left. Axes
units are in mas, and the (0,0) position corresponds to the source peak in each
image. The convolving beam, plotted in the lower left corner, has a size of
3.4$\times$1.2~mas in PA of $0^{\circ}$. The first contour corresponds to 5
times the rms noise of the image (0.08 and 0.11~mJy~beam$^{-1}$ for run~A
and B, respectively), while consecutive ones scale with $2^{1/2}$. The dates
and orbital phases are quoted in the images. There is extended radio emission
that appears nearly symmetric for run~A and clearly asymmetric for run~B, with
a small change of $\sim12^{\circ}$ in its position angle.
\label{fig:vlba}}
\end{figure}

We show the final VLBA+phased VLA self-calibrated images in
Fig.~\ref{fig:vlba}. The image obtained for run~A displays a central core and
bipolar and nearly symmetric extended emission with PA~$\simeq116\pm2^{\circ}$,
with the brightest component towards the south-east. The image is similar to
the one obtained with the same array in 1999 May 8, corresponding to orbital
phases 0.12--0.15, which showed a slightly more asymmetric extended emission in
PA$\sim125^{\circ}$ (see \cite{paredes00}). In contrast, the image obtained for
run~B displays a core and a bipolar but clearly asymmetric structure with PA
$\simeq128\pm2^{\circ}$ and the brightest component towards the north-west. The
total and peak flux densities fitted to the data are shown in
Table~\ref{tab:parameters}. See also \cite{ribo08}.

We splitted each run into 4-hour data sets, and self-calibrated each block. No
significant morphological differences are measured between the two halves in
any of the two runs. The peak position of the component SE1 with respect to
Core1 is stable in 4 hours within the errors ($\sigma_{\alpha}^{\rm
A}=0.31$~mas, $\sigma_{\delta}^{\rm A}=0.62$~mas). For run~B the distance
between Core2 and NW2 is also stable in 4 hours within the errors
($\sigma_{\alpha}^{\rm B}=0.55$~mas, $\sigma_{\delta}^{\rm B}=0.25$~mas).

\begin{table}[t] 
\begin{center}
\caption{Parameters of the Gaussian components fitted to the data. Columns~3
and 4 list the peak and integrated flux densities of each component. Columns~5
to 8 list the polar and Cartesian coordinates of the components with respect to
the peak position. The PA is positive from north to east.}
\label{tab:parameters} 
\begin{tabular}{l@{~~}l@{~}r@{ $\pm$ }l@{~~}r@{ $\pm$ }l@{~~}r@{ $\pm$ }l@{~~}r@{ $\pm$ }l@{~~}r@{ $\pm$ }l@{~~}r@{ $\pm$ }l}
\hline
\hline
Run & Comp. & \multicolumn{2}{l}{Peak $S_{\rm 5~GHz}$} & \multicolumn{2}{c}{$S_{\rm 5~GHz}$} & \multicolumn{2}{c}{$r$}   & \multicolumn{2}{c}{PA}         & \multicolumn{2}{c}{$\Delta\alpha$} & \multicolumn{2}{c}{$\Delta\delta$}  \\
    &       & \multicolumn{2}{l}{[mJy~beam$^{-1}$]}    & \multicolumn{2}{c}{[mJy]}           & \multicolumn{2}{c}{[mas]} & \multicolumn{2}{c}{[$^{\circ}$]} & \multicolumn{2}{c}{[mas]}          & \multicolumn{2}{c}{[mas]}           \\
\hline
A   & Core1 & 10.54 & 0.08 & 20.0 & 0.2 & \multicolumn{2}{c}{---} & \multicolumn{2}{c}{---} & \multicolumn{2}{c}{---} & \multicolumn{2}{c}{---}  \\
    & SE1   &  1.11 & 0.08 &  2.6 & 0.2 & 3.67  & 0.08 & 115.9   & 1.7 &    3.30 & 0.07 &  $-$1.60 & 0.12  \\
    & NW1   &  0.88 & 0.08 &  1.5 & 0.2 & 3.29  & 0.09 & $-$63   & 2   & $-$2.92 & 0.08 &     1.52 & 0.14  \\
\hline
B   & Core2 & 10.45 & 0.11 & 17.6 & 0.3 & \multicolumn{2}{c}{---} &\multicolumn{2}{c}{---} &\multicolumn{2}{c}{---} &\multicolumn{2}{c}{---}  \\
    & SE2   &  0.75 & 0.11 &  1.8 & 0.4 & 2.8~~ & 0.2  &     129 & 5   &    2.17 & 0.13 & $-$1.8~~ & 0.3   \\
    & NW2   &  2.22 & 0.11 &  3.9 & 0.3 & 2.94  & 0.06 & $-$52.2 & 1.4 & $-$2.32 & 0.04 &     1.80 & 0.09  \\
\hline
\end{tabular}
\end{center}
\end{table}

The measured position of LS~5039 for run~A is $\alpha_{\rm J2000.0}=18^{\rm h}
26^{\rm m} 15\rlap.^{\rm s}05653\pm$0.15~mas, and $\delta_{\rm
J2000.0}=-14^\circ 50^\prime 54\rlap.{''}2564\pm$1.5~mas. For checking purposes
we measured the position of data blocks of 4 and 1-hour lengths. The peak
position of LS~5039 appears to move  $\Delta\alpha=+0.1\pm0.1$~mas,
$\Delta\delta=+2.8\pm0.2$~mas between the two 4-hour blocks (slightly more in
the 1-hour blocks). The observed direction has a PA of $1.7\pm1.7^\circ$, which
is the same as the line joining the positions of the source and J1825$-$1718.
The expected error for differential astrometry is given by the separation $d$
in degrees from the phase-reference source and the offset $\Delta$ of its
correlated position, according to $\Delta\times(d/180)\times\pi$ (see
\cite{walker99}). Plugging our offset of 60.9~mas and a distance of
2$\rlap.^{\rm\circ}$47 we obtain an error of 2.6~mas, very similar to the
observed displacement of LS~5039. Similar results are obtained when splitting
the data from the astrometric check source J1837$-$1532. Therefore, these
secular motions appear to be purely instrumental. The errors assigned to the
coordinates of LS~5039 quoted above are half of the total secular motion
measured in 1-hour blocks.

The phase-referenced image of run~B reveals a double source with similar peak
flux densities of $3.6\pm0.1$~mJy~beam$^{-1}$, much lower than expected, and
are separated $2.5\pm0.1$~mas in PA of $87\pm5^\circ$. Moreover, 4- and 1-hour
blocks reveal a fading of the western component and a brightening of the
eastern one along the run, as well as a similar secular motion as in run~A.
Tropospheric errors, which affect the phase-referenced image and cannot be
accounted for, can easily split Core2 into the observed double source. The
precise position of the peak of LS~5039 cannot be measured in run~B. The
relative astrometry between runs in these observations is therefore not useful.

\section{Discussion} \label{discussion}

The observations of LS~5039 reported here, obtained with the VLBA on two runs
separated by 5 days, show a changing morphology at mas scales. In both runs
there is a core component with a constant flux density within errors, and
elongated emission with a PA that changes by $12\pm3^{\circ}$ between both
runs. The brightest component in run~A is towards south-east, and in run~B
towards north-west. The source is nearly symmetric in run~A and asymmetric in
run~B (see Fig.~\ref{fig:vlba}). The relative astrometry obtained with current
data is not considered in this discussion.

In the microquasar scenario, and assuming ballistic motions of adiabatically
expanding plasma clouds without shocks \cite{mirabel99}, the morphology of
run~A can be interpreted as a double-sided jet emanating from a central core
with the southeast component as the approaching one, whereas in run~B the
northwest component would be the approaching one. We can compute the projected
bulk velocity of the jets from the distance and flux asymmetry. The results are
compatible with a mildly relativistic jet with a projected bulk velocity of
0.1c. However, the distances from Core2 to the components NW2 and SE2 are very
similar and do not imply any significant relativistic motion.

The lack of proper motions (see upper limits in Sect.~\ref{results}) sets an
upper limit to the projected bulk velocities of the components. For the
measured flux asymmetries, the SE1 jet should be pointing at
$\theta<48^{\circ}$, and the NW2 jet at $\theta<45^{\circ}$. In this context, a
fast jet precession is needed to explain this behavior, and the PA of the jet
should vary considerably, in contrast to the small range covered by all
observed values at mas scale, between 115 and 140$^{\circ}$
\cite{paredes00,paredes02}. See discussion in \cite{ribo08}.

Alternatively, the morphology detected in run~B could be the result of a
discrete ejection where Core2 is the approaching component and NW2 the receding
one, while there is no radio emission at the origin of the ejection. However,
large X-ray and radio flux density variations are expected during discrete
ejections (see \cite{fender06} and references therein), while the peak and
total radio flux densities of LS~5039 are strikingly constant (see also
\cite{ribo99} and \cite{clark01}) and there is no evidence of an X-ray flare in
11.5 years of {\it RXTE}/ASM data. Although not yet explored in detail, the
morphology changes can be due to the interaction between the jets and a clumpy
or dense stellar wind \cite{perucho08}.

In the non-accreting pulsar scenario, the different morphologies we have
detected at different orbital phases could be due to the change of the relative
positions between the pulsar and the companion star along the orbit (see
details in \cite{dubus06}). Observations at different orbital phases have
always revealed a very similar PA for the extended emission, which would
correspond to an inclination of the orbit as seen nearly edge on (90$^\circ$).
On the other hand, the absence of X-ray eclipses places an upper limit of
$i\lesssim 75^{\circ}$ in this scenario. Therefore, these two restrictions
imply an inclination angle that should be close to the upper limit of
75$^{\circ}$. 

In conclusion, a simple and shockless microquasar scenario cannot easily
explain the observed changes in morphology. On the other hand, an
interpretation within the young non-accreting pulsar scenario requires the
inclination of the binary system to be very close to the upper limit imposed by
the absence of X-ray eclipses. Precise phase-referenced VLBI observations
covering a whole orbital cycle are necessary to trace possible periodic
displacements of the peak position, expected in this last scenario, and to
obtain morphological information along the orbit. These will ultimately reveal
the nature of the powering source in this gamma-ray binary.

\acknowledgments{
The NRAO is a facility of the National Science Foundation operated under
cooperative agreement by Associated Universities, Inc. J.M., M.R., J.M.P., and
J.M. acknowledge support by DGI of the Spanish Ministerio de Educaci\'on y
Ciencia (MEC) under grants AYA2007-68034-C03-01 and AYA2007-68034-C03-02 and
FEDER funds. J.M. was supported by MEC under fellowship BES-2008-004564. M.R.
acknowledges financial support from MEC and European Social Funds through a
\emph{Ram\'on y Cajal} fellowship. J.M. is also supported by Plan Andaluz de
Investigaci\'on of Junta de Andaluc\'{\i}a as research group FQM322. This work
has benefited from research funding from the European Community's sixth
Framework Programme under RadioNet R113CT 2003 5058187.
}


\begin{thebibliography}{99}

\bibitem{aharonian05a}
Aharonian, F., Akhperjanian, A.~G., Aye, K.-M., et~al.\
2005a, A\&A, 442, 1

\bibitem{aharonian05b}
Aharonian, F., Akhperjanian, A.~G., Aye, K.-M., et~al.\
2005b, Science, 309, 746

\bibitem{albert06}
Albert, J., Aliu, E., Anderhub, H., et~al.\
2006, Science, 312, 1771

\bibitem{albert07}
Albert, J., Aliu, E., Anderhub, H., et~al.\
2007, ApJ, 665, L51


\bibitem{bosch06}
Bosch-Ramon, V., Romero, G.~E., \& Paredes, J.~M.\
2006, A\&A, 447, 263

\bibitem{casares05}
Casares, J., Rib\'o, M., Ribas, I., et~al.\
2005, MNRAS, 364, 899

\bibitem{clark01}
Clark, J.~S., Reig, P., Goodwin, S.~P., et~al.\
2001, A\&A, 376, 476


\bibitem{dubus06}
Dubus, G.\
2006, A\&A, 456, 801

\bibitem{fender06}
Fender, R.~P.\
2006, Compact Stellar X-Ray Sources, ed.\ W.~H.~G. Lewin \& M. van der Klis (Cambridge University Press)

\bibitem{maraschi81}
Maraschi, L., \& Treves, A.\
1981, MNRAS, 194, 1P 

\bibitem{marti98}
Mart\'{\i}, J., Paredes, J.~M., \& Rib\'o, M.\
1998, A\&A, 338, L71

\bibitem{massi04}
Massi, M., Rib\'o, M., Paredes, J.~M., et~al.\
2004, A\&A, 414, L1

\bibitem{mirabel99}
Mirabel, I.~F., \& Rodr\'{\i}guez, L.~F.\
1999, ARA\&A, 37, 409

\bibitem{morris02}
Morris, D.~J., Hobbs, G., Lyne, A. G., et al.\ 
2002, MNRAS, 335, 275 

\bibitem{paredes00}
Paredes, J.~M., Mart{\'{\i}}, J., Rib{\'o}, M., \& Massi, M.\
2000, Science, 288, 2340 

\bibitem{paredes02}
Paredes, J.~M., Rib\'o, M., Ros, E., Mart\'{\i}, J., \& Massi, M.\
2002, A\&A, 393, L99 

\bibitem{paredes06}
Paredes, J.~M., Bosch-Ramon, V., \& Romero, G.~E.\
2006, A\&A, 451, 259

\bibitem{perucho08}
Perucho, M., \& Bosch-Ramon, V.\ 
2008, A\&A, 482, 917 

\bibitem{ribo99}
Rib\'o, M., Reig, P., Mart\'{\i}, J., \& Paredes, J.~M.\
1999, A\&A, 347, 518 

\bibitem{ribo08}
Rib{\'o}, M., Paredes, J.~M., Mold{\'o}n, J., Mart\'{\i}, J., \& Massi, M.\
2008, A\&A, 481, 17 

\bibitem{romero03}
Romero, G.~E., Torres, D.~F., Kaufman Bernad\'o, M.~M., \& Mirabel, I.~F.\
2003, A\&A, 410, L1 

\bibitem{sierpowska07} 
Sierpowska-Bartosik, A., \& Torres, D.~F.\
2007, ApJ, 671, L145

\bibitem{walker99}
Walker, R.~C.\
1999, ASP Conf. Series, 180, 433

\end{thebibliography}
\end{document}